\documentclass[journal]{IEEEtran}
\usepackage[T1]{fontenc}
\usepackage[latin9]{inputenc}
\usepackage{array}
\usepackage{mathtools}
\usepackage{algorithm2e}
\usepackage{amsmath,amssymb}
\usepackage{graphicx}
\usepackage[unicode=true,
 bookmarks=true,bookmarksnumbered=true,bookmarksopen=true,bookmarksopenlevel=1,
 breaklinks=false,pdfborder={0 0 1},backref=false,colorlinks=false]
 {hyperref}
\hypersetup{pdftitle={Your Title},
 pdfauthor={Your Name},
 pdfborderstyle=,pdfpagelayout=OneColumn,pdfnewwindow=true,pdfstartview=XYZ,plainpages=false}

\makeatletter

\providecommand{\tabularnewline}{\\}

\makeatother

\begin{document}
\title{Quantum Circuits for Stabilizer Error Correcting Codes: A Tutorial}
\author{Arijit Mondal and Keshab K. Parhi, {\em Fellow, IEEE}\\
 Email: \{monda109, parhi\}@umn.edu\\
 Department of Electrical and Computer Engineering, University of
Minnesota}
\maketitle
\begin{abstract}
Quantum computers have the potential to provide exponential speedups over their classical counterparts.
Quantum principles are being applied to
fields such as communications, information processing, and artificial intelligence to achieve quantum advantage.
However, quantum bits are extremely noisy and prone to decoherence. Thus, keeping the qubits
error free is extremely important toward reliable quantum
computing. Quantum error correcting codes have been studied for several decades and methods have been proposed to import classical
error correcting codes to the quantum domain. However, circuits for
such encoders and decoders haven't been explored in depth. This paper serves as a tutorial on designing and simulating quantum encoder and decoder circuits for stabilizer codes. We present encoding and decoding circuits for five-qubit code and Steane code, along with verification of these circuits using IBM Qiskit. We also provide nearest neighbour compliant encoder and decoder circuits for the five-qubit code.\end{abstract}


\section{Introduction}

Quantum computing is a rapidly-evolving technology which exploits
the fundamentals of quantum mechanics towards solving tasks which
are too complex for current classical computers. 
Quantum computers have the potential to achieve exponential speedups over their
classical counterparts \cite{arute, kim}.
In 1981, Feynman suggested that a quantum computer would have the power to
simulate systems which are not feasible for classical computers \cite{feynman-1,feynman-2}.
In 1994, Shor proposed a quantum algorithm to find the prime
factors of an integer in {\em polynomial} time \cite{shor-factorization}.
Grover proposed an algorithm which was able to search a particular
element in an unsorted database with a high probability, with significantly
higher efficiency than any known classical algorithm in 1996 \cite{grover}.
Subsequently, several quantum algorithms aimed at achieving better efficiencies than their classical counterparts were proposed. However, practical realization of these algorithms requires quantum computers,
which are slowly evolving. IBM recently demonstrated a 433 qubit quantum
computer \cite{osprey}, and expects to deliver a quantum computer having more than a thousand qubits within one year. The path towards a powerful quantum computer which can perform Shor's factorization or Grover's search
algorithm may not be a distant reality. However, a fundamental issue
needs to addressed. As we pack more number of qubits into quantum processors, we need to have a reliable method of processing to mitigate noise and quantum decoherence.

The phenomenon through which quantum mechanical systems attain interference
among each other is known as quantum coherence. Quantum coherence
is essential to perform quantum computations on quantum information. However, quantum systems are inherently susceptible to noise and decoherence which necessitates building fault tolerant systems which can overcome
noise and decoherence. Thus, quantum error correcting codes (ECCs) become a necessity for quantum computing systems. There were various challenges in the way of designing a quantum ECC framework. It is
well known that measurement destroys superpositions in any quantum
system. Additionally, since the quantum errors are continuous in nature, the
design of an ECC for quantum systems was difficult. To make things more complicated, the no-go theorems in the quantum realm make it challenging to design an ECC system analogous to classical domain \cite{nielsen,wootters,dieks,pati,barnum}. Quantum ECCs were
believed to be impossible till 1995, when Shor demonstrated a 9-qubit
ECC which was capable of correcting a single qubit error for the first
time \cite{shor}. In 1996, Gottesman proposed a stabilizer framework
which was widely used for the construction of quantum ECCs from classical
ECCs \cite{gottesman,gottesman-thesis}. Calderbank-Shor-Steane (CSS)
codes were proposed independently by Calderbank-Shor \cite{calderbank}
and Steane \cite{steane}. These codes were used to derive quantum
codes from binary classical linear codes which satisfy a dual-containing
criterion. The necessary and sufficient conditions for a quantum ECC
to be able to recover from a set of errors were given 
in \cite{knill}. Topological quantum codes like toric code were constructed for applications on quantum circuits arranged in a torus \cite{kitaev}. Subsequently, surface codes were introduced using stabilizer formalism in \cite{kitaev2}.

\begin{figure}
\begin{centering}
\includegraphics[scale=0.195]{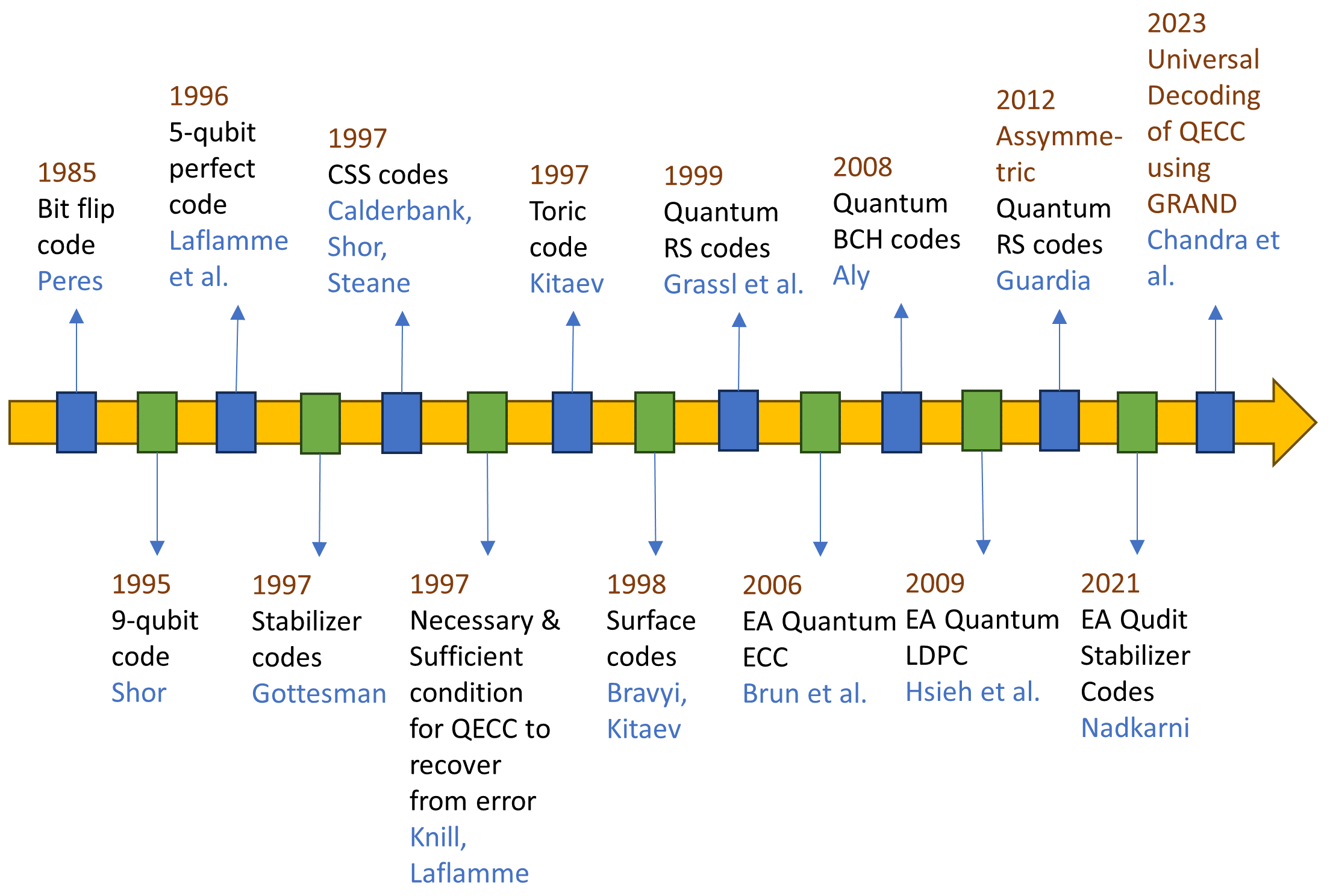}
\par\end{centering}
\caption{Chronological list of some of the primary advances in quantum ECC. \label{fig:chronology}}
\end{figure}

Pre-shared entangled qubits were proposed toward constructing stabilizer codes over non-Abelian groups in \cite{brun}.
This is done by extending the non-Abelian group into an Abelian group
by using extended operators which commute with each other. These entanglement-assisted
(EA) stabilizer codes contain qubits over the extended operators which
are assumed to be at the receiver end throughout, and entangled with
the transmitted set of qubits. It was later shown that EA stabilizer
codes increase the error correcting capability of quantum ECCs \cite{lai-brun}.
The advantage of the stabilizer framework lies in its ability to
construct quantum ECCs from any classical binary ECC. The optimal
number of pre-shared entangled qubits required for an EA stabilizer
code was expressed analytically, along with an encoding
procedure in \cite{wilde}. Quantum analog of classical low-density parity-check (LDPC) codes were constructed using quasi-cyclic
binary LDPC codes in \cite{hsieh}. Algebraic codes like 
Reed Solomon (RS) codes were also explored in the quantum domain by the authors in \cite{grassl}, \cite{aly}, and \cite{guardia} using self-orthogonal classical RS codes. Purely quantum polar codes based on recursive channel combining and splitting construction were studied in \cite{dupuis}. EA stabilizer codes were extended to qudit systems in \cite{nadkarni}. Recently, a universal decoding scheme was conceived for quantum stabilizer codes (QSCs) by adapting 'guessing random additive noise decoding' (GRAND) philosophy from classical domain codes \cite{chandra}. However, it becomes necessary to design actual encoder and decoder
circuits for these quantum ECCs, so that reliable quantum computing systems can be built. The CSS framework is particularly interesting due to its simplicity. It provides a method for importing any classical
ECC into the quantum domain, as long as the dual-containing criterion is satisfied. A chronological list of some of the primary advances in quantum ECCs is shown in Fig. \ref{fig:chronology}.

The contributions of this paper are as follows. First, we revisit the systematic method for construction of encoder and decoder circuits for stabilizer codes. We identify and analyze the key concepts for the construction of an encoder for stabilizer codes, demonstrated in \cite{gottesman-thesis} through a five-qubit code \cite{bennett,laflamme}. The concepts are then used to formulate an algorithm for the construction of an encoder circuit for a general stabilizer code. For the decoder design, we use a syndrome measurement circuit, and depending on the measured syndromes, we may apply the appropriate error
correction using suitable Pauli gates. Second, we present encoder and decoder
circuits for two stabilizer codes: the five-qubit code and the Steane code. Third, we provide nearest neighbour compliant (NNC) circuits for the encoding and syndrome measurement of five-qubit code \cite{ding}.
Fourth, we present simulation results using IBM Qiskit \cite{qiskit} and verify the circuits.  

The rest of the paper is organized as follows. Section II presents
a brief description of quantum gates and quantum circuits. Section III reviews Shor's 9-qubit code \cite{shor}, stabilizer formalism, and CSS codes. In Section IV, we provide a systematic method of construction of the encoder and decoder circuits for a general stabilizer code. Using the above knowledge, we present design of encoding and syndrome measurement circuits for the five-qubit code and Steane code in Sections V and VI, respectively. In Section VII, we provide nearest neighbour compliant circuits for the five-qubit code. We discuss the results and comparisons in Section VIII, followed by conclusions in Section IX.

\section{Pauli matrices and quantum gates}

In two-level quantum systems, the two-dimensional unit of quantum
information is called a quantum bit (qubit). The state of a qubit
is represented by $|\psi\rangle=a|0\rangle+b|1\rangle,$ where $a,b\in \mathbb{C}$
and $|a|^{2}+|b|^{2}=1$. $|0\rangle$ and $|1\rangle$ are basis
states of the state space. The evolution of a quantum mechanical system
is fully described by a unitary transformation. State $|\psi_{1}\rangle$
of a quantum system at time $t_{1}$ is related to $t_{2}$ by a unitary
operator $U$ that depends only on the time instances $t_{1}$ and
$t_{2}$, i.e., $|\psi_{2}\rangle=U|\psi_{1}\rangle$. The unitary
operators or matrices which act on the qubit belong to $\mathbb{C}^{2\times2}$.
We have a Pauli group which represents the unitary matrices given
by

\begin{equation}\label{eq:Pauli}
\Pi=\{\pm I_{2},\pm iI_{2},\pm X,\pm iX,\pm Y,\pm iY,\pm Z,\pm iZ\}
\end{equation}

where $I_{2}=\left[\begin{array}{cc}
1 & 0\\
0 & 1
\end{array}\right],\,X=\left[\begin{array}{cc}
0 & 1\\
1 & 0
\end{array}\right],\,Y=\left[\begin{array}{cc}
0 & -i\\
i & 0
\end{array}\right],\,Z=\left[\begin{array}{cc}
1 & 0\\
0 & -1
\end{array}\right]$.

A quantum circuit consists of an initial set of qubits as inputs which
evolve through time to a final state, comprising of
the outputs of the quantum circuit. Quantum states evolve through
unitary operations which are represented by quantum gates. Quantum
gates can be single qubit gates which act on a single qubit, or they
can be multiple qubit gates which act on multi-qubit states to produce
a new multi-qubit state. The single qubit gates include the bit flip gate
$X$, phase flip gate $Z$, Hadamard gate $H$, $Y$ gate, and the
phase gate $S$. The unitary operations related to the single qubit
gates are as follows:

\begin{align}
X & =\left[\begin{array}{cc}
0 & 1\\
1 & 0
\end{array}\right],Z=\left[\begin{array}{cc}
1 & 0\\
0 & -1
\end{array}\right],H=\frac{1}{\sqrt{2}}\left[\begin{array}{cc}
1 & 1\\
1 & -1
\end{array}\right],\\
Y & =\left[\begin{array}{cc}
0 & -i\\
i & 0
\end{array}\right],S=\left[\begin{array}{cc}
1 & 0\\
0 & i
\end{array}\right]\nonumber 
\end{align}

The multi-qubit gates include controlled-$X$ (CNOT), controlled-$Z$ (CZ), controlled-$Y$ gates, and the CCNOT (Toffoli gate). They act on 2-qubit pr 3-qubit states and are given by the
following unitary transformations:

\begin{equation}
CNOT=\left[\begin{array}{cccc}
1 & 0 & 0 & 0\\
0 & 1 & 0 & 0\\
0 & 0 & 0 & 1\\
0 & 0 & 1 & 0
\end{array}\right], CZ=\left[\begin{array}{cccc}
1 & 0 & 0 & 0\\
0 & 1 & 0 & 0\\
0 & 0 & 1 & 0\\
0 & 0 & 0 & -1
\end{array}\right],
\end{equation}

\begin{equation}
CY=\left[\begin{array}{cccc}
1 & 0 & 0 & 0\\
0 & 1 & 0 & 0\\
0 & 0 & 0 & -i\\
0 & 0 & i & 0
\end{array}\right]
\end{equation}

\begin{equation}
CCNOT=\left[\begin{array}{cccccccc}
1 & 0 & 0 & 0 & 0 & 0 & 0 & 0\\
0 & 1 & 0 & 0 & 0 & 0 & 0 & 0\\
0 & 0 & 1 & 0 & 0 & 0 & 0 & 0\\
0 & 0 & 0 & 1 & 0 & 0 & 0 & 0\\
0 & 0 & 0 & 0 & 1 & 0 & 0 & 0\\
0 & 0 & 0 & 0 & 0 & 1 & 0 & 0\\
0 & 0 & 0 & 0 & 0 & 0 & 0 & 1\\
0 & 0 & 0 & 0 & 0 & 0 & 1 & 0
\end{array}\right]
\end{equation}

\begin{figure}
\begin{centering}
\includegraphics[scale=0.5]{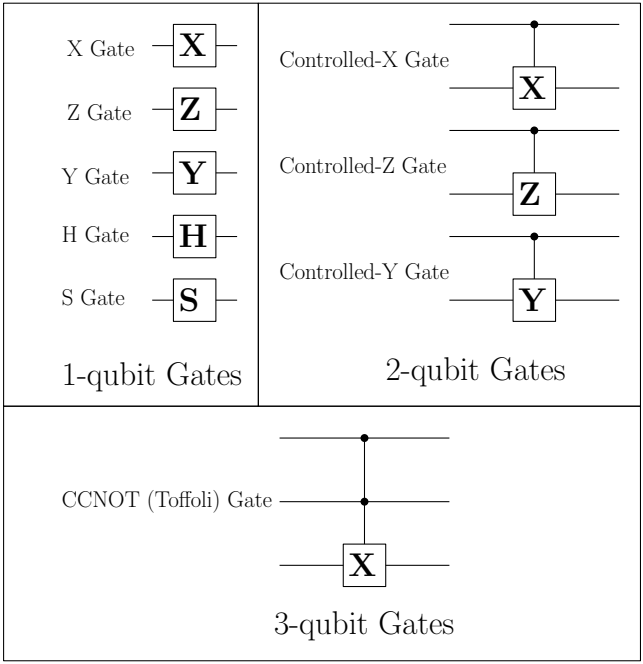}
\par\end{centering}
\caption{Symbolic representations of various 1-qubit and 2-qubit gates. \label{fig:gates}}
\end{figure}

Symbolic representations of various 1-qubit, 2-qubit, and 3-qubit gates are shown in Fig. \ref{fig:gates}.

\section{Shor's 9-qubit ECC, stabilizer formalism, and CSS codes}

Shor's 9-qubit code was the first ever quantum ECC capable
of correcting a single qubit error \cite{shor}. Gottesman proposed a general methodology to construct quantum
ECCs \cite{gottesman-thesis}. This method is known as the stabilizer construction and the
codes thus generated are known as stabilizer codes. Calderbank-Shor \cite{calderbank}
and Steane \cite{steane} proposed a method to derive quantum
codes from binary classical linear codes which satisfy a dual-containing
criterion. We will discuss the above in detail in this section.

\subsection{Shor's 9-qubit Quantum ECC}

Shor's 9-qubit code consists of a combination of 3-qubit bit flip and 3-qubit phase flip codes. First, we will provide a brief description of the working of these 3-qubit codes. From classical ECCs, we know about repetition codes. For a rate 1/3 repetition code, $0$ is transmitted as $000$ and $1$ is transmitted as $111$. The redundancies ensure that if a single error has occurred, a majority detector can detect and correct the error. Analogous to repetition code, we have a 3-qubit bit flip code and a 3-qubit phase flip code. However, due to the no-cloning theorem in quantum domain, we cannot create copies of a certain qubit state. Next, we will describe how this limitation is overcome towards the design of the 3-qubit bit flip and phase flip codes.

\subsubsection{3-qubit bit flip code}

\begin{figure}
\begin{centering}
\includegraphics[scale=0.6]{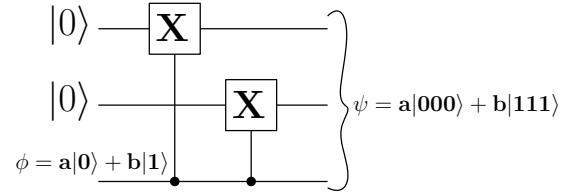}
\par\end{centering}
\caption{3-qubit bit flip encoder. \label{fig:single-bit-flip}}
\end{figure}

We can design a 3-qubit quantum code \cite{peres} capable of correcting a single bit flip error as shown in Fig. \ref{fig:single-bit-flip}. Two ancilla qubits are initialized to $|0\rangle$ analogous to redundant bits in a 3-bit repetition code. A single qubit is thus encoded into a 3-qubit state.
The basis states $|0\rangle$ and $|1\rangle$ are encoded using encoding
as shown below:
\begin{equation}
|0\rangle\xrightarrow{CNOT(3,2)CNOT(3,1)}|000\rangle,
\end{equation}
\begin{equation}
|1\rangle\xrightarrow{CNOT(3,2)CNOT(3,1)}|111\rangle    
\end{equation}
For an arbitrary normalized state $|\phi\rangle=a|0\rangle+b|1\rangle$,
where $a,b\in\mathbb{C}$, the encoding operation results in the state $|\psi\rangle$ given by
\begin{equation}
|\psi\rangle=a|000\rangle+b|111\rangle    
\end{equation}
The notation $CNOT(x,y)$ implies a $CNOT$ gate acting on qubits indexed $x$ and $y$, with $x$ as control and $y$ as target qubit. The qubits are numbered from top to bottom. It should be noted that in CNOT(3,2)CNOT(3,1), the rightmost operation happens first and the leftmost operation is performed last. 

\begin{figure}
\begin{centering}
\includegraphics[scale=0.6]{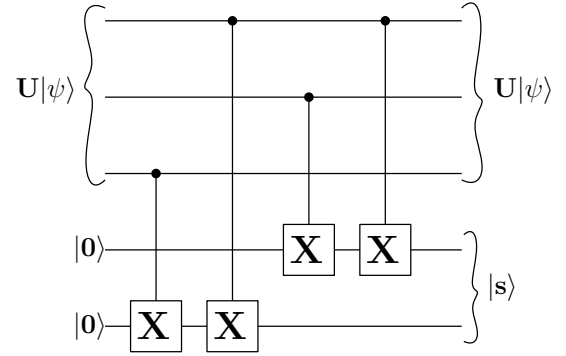}
\par\end{centering}
\caption{3-qubit bit flip syndrome computation. \label{fig:single-bit-flip-receiver}}
\end{figure}

The syndrome computation circuit is shown in Fig. \ref{fig:single-bit-flip-receiver}. Two ancilla qubits initialized to $|0\rangle$ are used to compute the syndrome. We perform the operation CNOT(1,4)CNOT(2,4)CNOT(1,5)CNOT(3,5)
on the state $\mathcal{U}|\psi\rangle|00\rangle$ to obtain $\mathcal{U}|\psi\rangle|s\rangle$
as shown in Fig. \ref{fig:single-bit-flip-receiver}. The two qubit syndrome state is given by $|s\rangle$.

Let's take an example to demonstrate the syndrome detection. Let the error be $\mathcal{U}=I_{2}\otimes I_{2}\otimes X$,
leading to the erroneous state $\mathcal{U}|\psi\rangle=a|001\rangle+b|110\rangle$. Performing the operation CNOT(1,4)CNOT(2,4)CNOT(1,5)CNOT(3,5)
on $\mathcal{U}|\psi\rangle|00\rangle$, we have \cite{nadkarni},

\begin{align*}
 & (CNOT(1,4)CNOT(2,4)CNOT(1,5)CNOT(3,5))\\
 & \mathcal{U}|\psi\rangle|00\rangle\\
= & (CNOT(1,4)CNOT(2,4)CNOT(1,5)CNOT(3,5))\\
& (a|001\rangle+b|110\rangle)|00\rangle\\
= & (CNOT(1,4)CNOT(2,4)CNOT(1,5)CNOT(3,5))\\
& (a|00100\rangle+b|11000\rangle)\\
= & a|00101\rangle+b|11001\rangle\\
= & \mathcal{U}|\psi\rangle|01\rangle\\
= & \mathcal{U}|\psi\rangle|s\rangle
\end{align*}

Thus, the syndrome is $|s\rangle=|01\rangle.$ The syndromes
$|11\rangle,$ $|10\rangle$, and $|01\rangle$ correspond to errors
in the first, second and third qubits respectively. Here, since the syndrome is $|01\rangle$, the third
qubit is in error.

\subsubsection{3-qubit phase flip code}
A 3-qubit phase flip code encodes a single qubit into a 3-qubit state as shown in Fig. \ref{fig:single-phase-flip}. Basis states $|0\rangle$ and $|1\rangle$ are encoded
as shown below:

\begin{equation}
|0\rangle\xrightarrow{H^{\otimes3}CNOT(3,2)CNOT(3,1)}|+++\rangle,
\end{equation}

\begin{equation}
|1\rangle\xrightarrow{H^{\otimes3}CNOT(3,2)CNOT(3,1)}|---\rangle   
\end{equation}

where $|\pm\rangle=\frac{|0\rangle\pm|1\rangle}{\sqrt{2}}$. Any arbitrary normalized state $|\phi\rangle=a|0\rangle+b|1\rangle$ gets encoded to state $|\psi\rangle$ using the above encoding operation as

\begin{equation}
|\psi\rangle=a|+++\rangle+b|---\rangle
\end{equation}

The unitary operator
$H^{\otimes3}CNOT(3,2)CNOT(3,1)$ is applied to the message state $|\phi\rangle$
along with two ancilla bits
initially in the state $|0\rangle$ to perform the encoding. 

\begin{figure}
\begin{centering}
\includegraphics[scale=0.6]{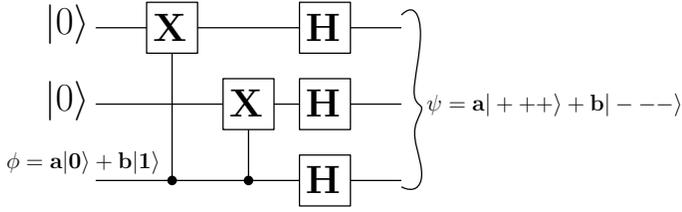}
\par\end{centering}
\caption{3-qubit phase flip encoder. \label{fig:single-phase-flip}}
\end{figure}

\begin{figure}
\begin{centering}
\includegraphics[scale=0.55]{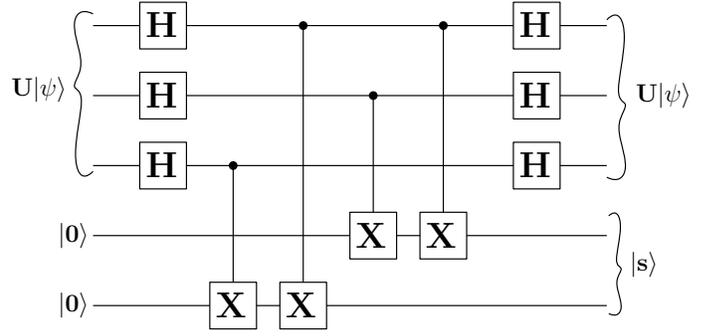}
\par\end{centering}
\caption{3-qubit phase flip syndrome computation. \label{fig:single-phase-flip-receiver}}
\end{figure}

The syndrome
detection circuit is shown in Fig. \ref{fig:single-phase-flip-receiver}. The syndrome is computed by performing the operation
$(H^{\otimes3}\otimes I_{2}^{\otimes2})(CNOT(1,4)CNOT(2,4)CNOT(1,5)CNOT(3,5))(H^{\otimes3}\otimes I_{2}^{\otimes2})$
on the state $\mathcal{U}|\psi\rangle|00\rangle$ to obtain $\mathcal{U}|\psi\rangle|s\rangle$,
where $|s\rangle$ is the two qubit syndrome state.

We now consider an example to demonstrate the syndrome detection. Let the error be $Z\otimes I_{2}\otimes I_{2}$. Thus, the erroneous
state is $U|\psi\rangle=a|-++\rangle+b|+--\rangle$. The first step $(H^{\otimes3}\otimes I_{2}^{\otimes2})$
converts $U|\psi\rangle|00\rangle$ to $U_{1}|\psi\rangle|00\rangle=a|10000\rangle+b|01100\rangle$. Next, the operation $(CNOT(1,4)CNOT(2,4)CNOT(1,5)CNOT(3,5))$
converts $U_{1}|\psi\rangle|00\rangle$ to $U_{2}|\psi\rangle|s\rangle$
as follows:

\begin{align*}
 & (CNOT(1,4)CNOT(2,4)CNOT(1,5)CNOT(3,5))\\
 & U_{1}|\psi\rangle|00\rangle\\
= & (CNOT(1,4)CNOT(2,4)CNOT(1,5)CNOT(3,5))\\
& (a|10000\rangle+b|01100\rangle)\\
= & a|10011\rangle+b|01111\rangle\\
= & a|100\rangle|11\rangle+b|011\rangle|11\rangle\\
= & (a|100\rangle+b|011\rangle)|11\rangle
\end{align*}

Next, the operation $(H^{\otimes3}\otimes I_{2}^{\otimes2})$
converts $(a|100\rangle+b|011\rangle)|11\rangle$ to $(a|-++\rangle+b|+--\rangle)|11\rangle=U|\psi\rangle|11\rangle=U|\psi\rangle|s\rangle$. Thus, the syndrome is $|s\rangle=|11\rangle$. The syndromes
$|11\rangle,$ $|10\rangle$, and $|01\rangle$ correspond to phase
errors in the first, second and third qubits respectively. Here, since the syndrome is $|11\rangle$, the first
qubit has a phase error.

The 3-qubit bit flip code is good at correcting a single bit flip. However, it cannot correct phase errors. It is in fact more prone to phase flip errors since phase flips in any of the qubits are indistinguishable from each other. Similarly, the 3-qubit phase flip code cannot correct bit flip errors. Hence, it was believed for a long time that a general quantum ECC capable of correcting both type of errors was not feasible, until Shor \cite{shor} proposed a 9 qubit code capable of correcting a bit flip and a phase flip simultaneously. We will discuss the encoding and decoding of this 9-qubit code in the following paragraphs.

\begin{figure}
\begin{centering}
\includegraphics[scale=0.45]{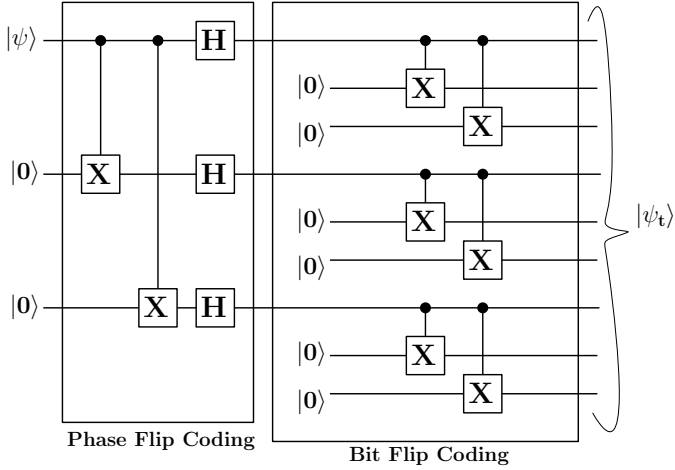}
\par\end{centering}
\caption{Encoder for Shor's 9-qubit code. \label{fig:shor-encoder}}
\end{figure}

The encoding circuit for the 9-qubit code is shown in Fig. \ref{fig:shor-encoder}. The encoding process can be broken into the following steps:

\textbf{Step 1:} Phase flip coding: After applying the CNOT gates we have the following state

\begin{equation}
|\psi_1\rangle=a|000\rangle+b|111\rangle
\end{equation}

Next, we have three Hadamard gates resulting in the state

\begin{align*}
|\psi_2\rangle & =a\left[\left(\frac{|0\rangle+|1\rangle}{\sqrt{2}}\right)\left(\frac{|0\rangle+|1\rangle}{\sqrt{2}}\right)\left(\frac{|0\rangle+|1\rangle}{\sqrt{2}}\right)\right]\\
 & +b\left[\left(\frac{|0\rangle-|1\rangle}{\sqrt{2}}\right)\left(\frac{|0\rangle-|1\rangle}{\sqrt{2}}\right)\left(\frac{|0\rangle-|1\rangle}{\sqrt{2}}\right)\right]
\end{align*}

\textbf{Step 2:} Bit flip coding: After adding the ancillas, we have the state

\begin{align*}
 |\psi_3\rangle =
 & \frac{a}{2\sqrt{2}}[\left((|0\rangle+|1\rangle)|00\rangle\right)\left((|0\rangle+|1\rangle)|00\rangle\right)\\
 & \left((|0\rangle+|1\rangle)|00\rangle\right)] + \frac{b}{2\sqrt{2}}[\left((|0\rangle-|1\rangle)|00\rangle\right)\\
 & \left((|0\rangle-|1\rangle)|00\rangle\right)\left((|0\rangle-|1\rangle)|00\rangle\right)]
\end{align*}

Next the CNOT gates are applied to achieve the encoded state 

\begin{align*}
|\psi_t\rangle = 
 & \frac{a}{2\sqrt{2}}\left[\left(|000\rangle+|111\rangle\right)\left(|000\rangle+|111\rangle\right)\left(|000\rangle+|111\rangle\right)\right]\\
 & +\frac{b}{2\sqrt{2}}\left[\left(|000\rangle-|111\rangle\right)\left(|000\rangle-|111\rangle\right)\left(|000\rangle-|111\rangle\right)\right]
\end{align*}

\begin{figure}
\begin{centering}
\includegraphics[scale=0.4]{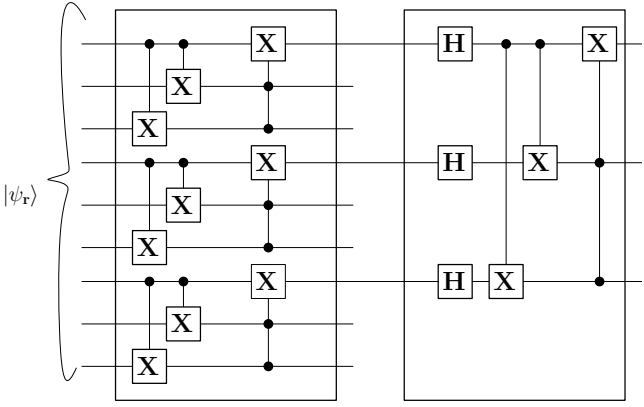}
\par\end{centering}
\caption{Decoder for Shor's 9-qubit code. \label{fig:shor-decoder}}
\end{figure}

The decoding circuit for the 9-qubit code is shown in Fig. \ref{fig:shor-decoder}. Let us assume that there is a bit and a phase flip on
the $4^{\mathrm{th}}$ qubit. Thus the combined state of the received
qubits can be represented as:

\begin{align*}
|\psi_{r}\rangle & =a\Bigg[\left(\frac{|000\rangle+|111\rangle}{\sqrt{2}}\right)\left(\frac{|100\rangle-|011\rangle}{\sqrt{2}}\right)\\
& \left(\frac{|000\rangle+|111\rangle}{\sqrt{2}}\right)\Bigg] +b\Bigg[\left(\frac{|000\rangle-|111\rangle}{\sqrt{2}}\right)\\
& \left(\frac{|100\rangle+|011\rangle}{\sqrt{2}}\right)\left(\frac{|000\rangle-|111\rangle}{\sqrt{2}}\right)\Bigg]
\end{align*}

The evolution of states for the decoding can be described in the following steps:

\textbf{Step 1:} After the application of the first two CNOT gates, we have

\begin{align*}
|\psi_{s_{1}}\rangle & =a\Bigg[\left(\frac{|000\rangle+|100\rangle}{\sqrt{2}}\right)\left(\frac{|111\rangle-|011\rangle}{\sqrt{2}}\right)\\
&\left(\frac{|000\rangle+|100\rangle}{\sqrt{2}}\right)\Bigg]+b\Bigg[\left(\frac{|000\rangle-|100\rangle}{\sqrt{2}}\right)\\
&\left(\frac{|111\rangle+|011\rangle}{\sqrt{2}}\right)\left(\frac{|000\rangle-|100\rangle}{\sqrt{2}}\right)\Bigg]
\end{align*}

\textbf{Step 2:} Next, the CCNOT (Toffoli) gates are applied resulting in the state

\begin{align*}
|\psi_{s_{2}}\rangle =& a\Bigg[\left(\frac{|000\rangle+|100\rangle}{\sqrt{2}}\right)\left(\frac{|011\rangle-|111\rangle}
{\sqrt{2}}\right)\\
 & \left(\frac{|000\rangle+|100\rangle}{\sqrt{2}}\right)\Bigg]
 +b\Bigg[\left(\frac{|000\rangle-|100\rangle}{\sqrt{2}}\right)\\
 & \left(\frac{|011\rangle+|111\rangle}{\sqrt{2}}\right)\left(\frac{|000\rangle-|100\rangle}{\sqrt{2}}\right)\Bigg]\\
&=a\Bigg[\left(\frac{(|0\rangle+|1\rangle)|00\rangle}{\sqrt{2}}\right)\left(\frac{(|0\rangle-|1\rangle)|11\rangle}{\sqrt{2}}\right)\\
& \left(\frac{(|0\rangle+|1\rangle)|00\rangle}{\sqrt{2}}\right)\Bigg]+b\Bigg[\left(\frac{(|0\rangle-|1\rangle)|00\rangle}{\sqrt{2}}\right)\\
& \left(\frac{(|0\rangle+|1\rangle)|11\rangle}{\sqrt{2}}\right)\left(\frac{(|0\rangle-|1\rangle)|00\rangle}{\sqrt{2}}\right)\Bigg]
\end{align*}

\textbf{Step 3:} Applying Hadamard
gate on $1^{\mathrm{st}}$, $4^{\mathrm{th}}$, and $7^{\mathrm{th}}$
qubits, we have 

\begin{equation}
|\psi_{s_{3}}\rangle=a|0\rangle_{1}|1\rangle_{4}|0\rangle_{7}+b|1\rangle_{1}|0\rangle_{4}|1\rangle_{7}
\end{equation}

\textbf{Step 4:} Next, applying CNOT gates, we have

\begin{equation}
|\psi_{s_{4}}\rangle=a|0\rangle_{1}|1\rangle_{4}|0\rangle_{7}+b|1\rangle_{1}|1\rangle_{4}|0\rangle_{7}
\end{equation}

\textbf{Step 5:} Finally, applying the CCNOT (Tofolli) gates, the state is 

\begin{align*}
|\psi_{s_{5}}\rangle =& |0\rangle_{1}|1\rangle_{4}|0\rangle_{7}+b|1\rangle_{1}|1\rangle_{4}|0\rangle_{7}\\
=&(a|0\rangle+b|1\rangle)|1\rangle|0\rangle
\end{align*}

As we can see, the first qubit is restored to the $a|0\rangle+b|1\rangle$ state. This is true independent of the index of the qubit on which the error has occurred. 

\subsection{Shor's 9-qubit code in stabilizer framework}

Now, we analyze the 9-qubit code and try to reason why it works, and then we generalize it toward a systematic method of error correction using the idea used in the 9-qubit code. From Fig. \ref{fig:shor-decoder}, we observe that for detecting bit flips in each group of three, we compare the first and third qubit, followed by the first two qubits. A correctly encoded state has the property that the first two qubits have even parity. Equivalently, a codeword
is a $+1$ eigen vector of $ZZI$, and a state with
an error on first or second qubit is
a $-1$ eigenvector of $ZZI$. Similarly, first
and third qubit should have even parity. Thus a codeword is also $+1$ eigenvector of $ZIZ$.

For detecting phase errors, we
compare the signs of first and second blocks of three, and the signs
of first and third blocks of three. Thus, a correctly encoded codeword is a $+1$
eigenvector of $XXXXXXIII$ and $XXXIIIXXX$. Thus, to correct the code, we need to measure
the eigenvalues of the eight
operators as shown in the following table.

\begin{tabular}{c|ccccccccc}
{$M_{1}$} & {$Z$} & {$Z$} & {$I$} & {$I$} & {$I$} & {$I$} & {$I$} & {$I$} & {$I$}
\tabularnewline
{$M_{2}$} & {$Z$} & {$I$} & {$Z$} & {$I$} & {$I$} & {$I$} & {$I$} & {$I$} & {$I$}
\tabularnewline
{$M_{3}$} & {$I$} & {$I$} & {$I$} & {$Z$} & {$Z$} & {$I$} & {$I$} & {$I$} & {$I$}
\tabularnewline
{$M_{4}$} & {$I$} & {$I$} & {$I$} & {$Z$} & {$I$} & {$Z$} & {$I$} & {$I$} & {$I$}
\tabularnewline
{$M_{5}$} & {$I$} & {$I$} & {$I$} & {$I$} & {$I$} & {$I$} & {$Z$} & {$Z$} & {$I$}
\tabularnewline
{$M_{6}$} & {$I$} & {$I$} & {$I$} & {$I$} & {$I$} & {$I$} & {$Z$} & {$I$} & {$Z$}
\tabularnewline
{$M_{7}$} & {$X$} & {$X$} & {$X$} & {$X$} & {$X$} & {$X$} & {$I$} & {$I$} & {$I$}
\tabularnewline
{$M_{8}$} & {$X$} & {$X$} & {$X$} & {$I$} & {$I$} & {$I$} & {$X$} & {$X$} & {$X$}
\tabularnewline
\end{tabular}{\scriptsize\par}

The two valid codewords in Shor's code are eigenvectors of all these operators $M_{1}$ through $M_{8}$ with eigenvalues $+1$. These generate a group, the stabilizer of the code, which consists of all Pauli operators $M$ with the property that $M|\psi\rangle=|\psi\rangle$ for all
encoded states $|\psi\rangle$.

\subsection{Binary vector space representation for stabilizers} \label{sec-2-b}

The stabilizers can be written as binary vector spaces, which can be useful to bring connections with classical error correction theory \cite{gottesman-thesis}. For this, the stabilizers are written as a pair of $(n-k)\times n$ matrices. The rows correspond to the stabilizers and the columns correspond to the qubits. The first matrix has a $1$ wherever there is a $X$ or $Y$ in the corresponding stabilizer, and $0$  everywhere else. The second matrix has a $1$ wherever there is a $Z$ or $Y$ in the corresponding stabilizer and $0$ everywhere else. It is often more convenient to write the two matrices as a single $(n-k)\times 2n$ matrix with a vertical line separating the two.

\subsection{Stabilizer formalism}

An $[[n,k]]$ quantum code can be used
for quantum error correction, where $k$ logical qubits are encoded using
$n$ physical qubits, leading to a code rate of $k/n$ analogous to classical
error correction. It has $2^k$ basis codewords, and any linear combination of the basis codewords are also valid codewords. Let the space of valid codewords be denoted by $T$. If we consider the tensor product of Pauli operators (with possible overall factors of $\pm 1$ or $\pm i$) in equation \ref{eq:Pauli}, it forms a group $G$ under multiplication. The stabilizer $S$ is
an Abelian subgroup of $G$, such that the code space $T$ is the space of vectors fixed by $S$ \cite{gottesman,gottesman-thesis}. Stabilizer generators are a set of independent set of $n-k$ elements from the stabilizer group, in the sense that none of them is a product of any two other generators.

We know that the operators in the Pauli group act on single qubit
states which are represented by $2$-bit vectors. The operators in
$\Pi$ have eigen values $\pm1$, and either commute or anti-commute
with other elements in the group. The set $\Pi^{n}$ is given by the
$n$-fold tensor products of elements from the Pauli group $\Pi$
as shown below,

\begin{align}
\Pi^{n}= & \{e^{i\phi}A_{1}\otimes A_{2}\otimes\cdots\otimes A_{n}\nonumber \\
 & :\forall j\in\{1,2,\cdots,n\}A_{j}\in\Pi,\phi\in\{0,\pi/2,\pi,3\pi/2\}\}
\end{align}

The stabilizers form a group with elements $M$ such that $M|\psi\rangle=|\psi\rangle$.
The stabilizer is Abelian, i.e., every pair of elements in the stabilizer
group commute. This can be verified from the following observation.
If $M|\psi\rangle=|\psi\rangle$ and $N|\psi\rangle=|\psi\rangle$,
then $MN|\psi\rangle-NM|\psi\rangle=(MN-NM)|\psi\rangle=0$. Thus,
$MN-NM=0$ or $MN=NM$, showing that every pair of elements in the
stabilizer group commute.

Given an Abelian subgroup $S$ of $n$-fold Pauli operators, the code space is defined
as

\begin{equation}
T(S)=\{|\psi\rangle,s.t.\,M|\psi\rangle=|\psi\rangle,\forall M\in S\}
\end{equation}

Suppose $M\in S$ and Pauli operator $E$ anti-commutes with $M$.
Then, $M(E|\psi\rangle)=-EM|\psi\rangle=-E|\psi\rangle$. Thus, $E|\psi\rangle$
has eigenvalue $-1$ for $M$. Conversely, if Pauli operator $E$
commutes with $M$, $M(E|\psi\rangle)=EM|\psi\rangle=E|\psi\rangle$,
thus $E|\psi\rangle$ has eigenvalue $+1$ for $M$. Thus, eigenvalue
of an operator $M$ from a stabilizer group detects errors which anti-commute
with $M$.

Single qubit operators $X$, $Y$, and $Z$  commute with themselves while they anti-commute with each other. For two multiple qubit operators, we need to evaluate how many anti-commutations happen. If the number is odd, the operators anti-commute; else, they commute. 

\textbf{Examples:} 

\begin{itemize}
    \item $X$  commutes with $X$, and anti-commutes with $Y$ and $Z$.
    \item $X\otimes X \otimes Z$  commutes with $X\otimes Y \otimes X$ since there are two anti-commuting qubit positions, 2 and 3.
    \item $Y\otimes Z \otimes X$ anti-commutes with $Y\otimes X \otimes X$ , since there is a single anti-commutation at position 2.
\end{itemize}

\subsection{CSS framework}

The CSS framework \cite{calderbank,steane} is a method to construct quantum ECCs from
their classical counterparts. Given two classical codes $C_{1}[n,k_{1},d_{1}]$
and $C_{2}[n,k_{2},d_{2}]$ which satisfy the dual containing criterion
$C_{1}^{\perp}\subset C_{2}$, CSS framework can be used to construct
quantum codes from such codes.

The CSS codes form a class of stabilizer codes. From the classical
theory of error correction, let $H_{1}$ and $H_{2}$ be the check
matrices of the codes $C_{1}$ and $C_{2}$. Since $C_{1}^{\perp}\subset C_{2}$,
codewords of $C_{2}$ are basically the elements of $C_{1}^{\perp}$.
Hence, we have, $H_{2}H_{1}^{T}=0$. The check matrix of a CSS code
is given by:

\begin{equation}
H_{C_{1}C_{2}}=\left[\begin{array}{ccc}
\begin{array}{c}
H_{1}\\
0
\end{array} & \Bigg| & \begin{array}{c}
0\\
H_{2}
\end{array}\end{array}\right]
\end{equation}

\section{Systematic procedure for encoder design for a stabilizer code}\label{sec:systematic}

A systematic method for the design of an encoder for a stabilizer code was presented in \cite{gottesman-thesis}. The encoder circuit for the five-qubit code proposed in \cite{gottesman-thesis} had a few errors which were later addressed in an errata \cite{gottesman-errata}. Taking those into consideration and making slight modifications, the complete procedure for the design of an encoder circuit for a stabilizer code can be summarized as follows:

\textbf{Step 1}: The stabilizers are written in a matrix form using binary vector space formalism as mentioned in Section \ref{sec-2-b}. Let the parity check matrix thus obtained be $H_{q}$.

\textbf{Step 2}: Our aim is to bring $H_{q}$ to the standard form $H_{s}$ below:
\begin{equation} \label{eq:std-hs}
H_{s}=\left[\begin{array}{ccc}
\begin{array}{ccc}
I_{1} & A_{1} & A_{2}\\
0 & 0 & 0
\end{array} & \Bigg| & \begin{array}{ccc}
B & C_{1} & C_{2}\\
D & I_{2} & E
\end{array}\end{array}\right]
\end{equation}

where, $I_{1}$ and $B$ are $r\times r$ matrices. `$r$' is the rank of the $X$ portion of $H_{s}$. $A_{1}$ and $C_{1}$
are $r\times(n-k-r)$ matrices. $A_{2}$ and $C_{2}$ are $r\times k$
matrices. $D$ is a $(n-k-r)\times r$ matrix. $I_{2}$ is a $(n-k-r)\times(n-k-r)$
matrix. $E$ is a $(n-k-r)\times k$ matrix. $I_{1}$ and $I_{2}$
are identity matrices. 

$H_{q}$ is converted to standard form $H_{s}$ using Gaussian elimination \cite{gottesman-thesis}. The logical operators $\overline{X}$ and $\overline{Z}$ can be written as 
\begin{equation}
\overline{X}=\left[\begin{array}{ccc}
\begin{array}{ccc}
0 & U_{2} & U_{3}\end{array} & | & \begin{array}{ccc}
V_{1} & 0 & 0\end{array}\end{array}\right]
\end{equation}

\begin{equation}
\overline{Z}=\left[\begin{array}{ccc}
\begin{array}{ccc}
0 & 0 & 0\end{array} & | & \begin{array}{ccc}
V'_{1} & 0 & V'_{3}\end{array}\end{array}\right]
\end{equation}

where $U_{2}=E^T$, $U_{3}=I_{k\times k}$, $V_{1}=E^{T}C_1^{T}+C_2^{T}$, $V'_{1}=A_{2}^{T}$, and $V'_{3}=I_{k\times k}$.

Given the parity check matrix in standard form $H_{s}$ and $\overline{X}$, the encoding operation for a stabilizer code can be written as,

\begin{align}
|c_{1}c_{2}\cdots c_{k}\rangle= & \overline{X}_{1}^{c_{1}}\overline{X}_{2}^{c_{2}}\cdots \overline{X}_{k}^{c_{k}}\left(\sum_{M\in S}M\right)|00\cdots 0\rangle\\
= & \overline{X}_{1}^{c_{1}}\overline{X}_{2}^{c_{2}}\cdots \overline{X}_{k}^{c_{k}}(I+M_{1})(I+M_{2})\cdots\nonumber \\
 & (I+M_{n-k})|00\cdots 0\rangle.
\end{align}

There are a total of $n$ qubits. Place qubits initialized to $|0\rangle$ at qubit positions $i=1$ to $i=n-k$. Place the qubits to be encoded at positions $i=n-k+1$ to $i=n$.

We observe the following from $H_{s}$ and $\overline{X}$:

\begin{itemize}

\item We know that a particular logical operator $\overline{X}_i$ is applied only if the qubit at $i^{\mathrm{th}}$ position is $|1\rangle$. Thus, applying $\overline{X}_i$ controlled at $i^{\mathrm{th}}$ qubit encodes $\overline{X}_i$.

\item The $\overline{X}$ operators consist of products of only $Z$s for the first $r$ qubits. For the rest of the qubits, $\overline{X}$ consists of products of $X$'s only. We know that $Z$ acts trivially on $|0\rangle$. Since the first $r$ qubits are initialized to $|0\rangle$, we can ignore all the $Z$s in $\overline{X}$. 

\item The first $r$ generators in $H_{s}$ apply only a single bit flip to the first $r$ qubits. This implies that when $I+M_{i}$ is applied, the resulting state would be a sum of $|0\rangle$ and  $|1\rangle$ for the $i^{\mathrm{th}}$ qubit. This corresponds to applying $H$ gates to the first $r$ qubits, which puts each of the $r$ qubits in the state $\frac{1}{\sqrt{2}}(|0\rangle+|1\rangle)$.

\item If we apply $M_{i}$ conditioned on qubit $i$, it implies the application of $I+M_{i}$. The reason is as follows. When the control qubit $i$ is $|1\rangle$, $M_i$ needs to be applied to the combined qubit state. Since the qubit $i$ suffers from a bit flip $X$ only by the stabilizer $M_{i}$, it is already in flipped state when it is $|1\rangle$. Thus, only the rest of the operators in $M_{i}$ need to be applied. However, there would be an issue if $H_{s_{(i,i+n)}}$ is not $0$, i.e., there is a $Y$ instead of $X$. In that case, adding an $S$ gate after the $H$ gate resolves the issue.

\end{itemize}

\textbf{Step 3}: The observations in Step 2 can be used to devise an algorithm as shown in Algorithm \ref{alg:algo-1} to design the encoding circuit.

\RestyleAlgo{ruled}
\begin{algorithm}
\caption{Algorithm to generate encoding circuit from $H_{s}$ and $\overline{X}$ ($n$ = number of physical qubits, $k$ = number of logical qubits, $r$ = rank of $X$-portion of $H_{s}$).}\label{alg:algo-1}
\KwData{$H_{s}$, $\overline{X}$}
\KwResult{Encoding circuit}
 \For{$i=1$ \KwTo $k$}{ 
    \If{$\overline{X}_{i,i+n-k}==1$}{
        Place controlled dot at qubit $i+n-k$\
        }
    \For{$j=1$ \KwTo $n$}{
        \If{$i+n-k\neq j$}{
            \If{$\overline{X}_{i,j}==1$}{
                Place $X$ gate at qubit $j$ controlled at qubit $i+n-k$
            }
        }
    }
 }
 \For{$i=1$ \KwTo $r$}{ 
    \eIf{$H_{s_{(i,i+n)}}==0$}{
        Place $H$ gate followed by controlled dot at qubit $i$\
        }
        {
        Place $H$ gate followed by $S$ gate followed by controlled dot at qubit $i$\
        }
    \For{$j=1$ \KwTo $n$}{
        \If{$i\neq j$}
            {\If{$H_{s_{(i,j)}}==1$ \&\& $H_{i,j+n}==0$}{
                Place $X$ gate on qubit $j$ with control at qubit $i$\
                }
            \If{$H_{s_{(i,j)}}==0$ \&\&  $H_{i,j+n}==1$}{
                Place $Z$ gate on qubit $j$ with control at qubit $i$\
                }
            \If{$H_{s_{(i,j)}}==1$ \&\&  $H_{i,j+n}==1$}{
                Place $Y$ gate on qubit $j$ with control at qubit $i$\
                }
            }
        }
    }
    
\end{algorithm}

\section{5-qubit perfect code}

The five-qubit \cite{bennett,laflamme} ECC is the smallest quantum
ECC with the ability to correct a single qubit error. 
It is a cyclic code with a distance of $3$. The treatment
of the 5-qubit code in the stabilizer formalism was provided in \cite{gottesman-thesis}. We will revisit the concept in brief.
The stabilizers $M_1-M_4$  along with the logical $\bar{X}$ and $\bar{Z}$ operators for a 5-qubit ECC are given as follows:
\begin{center}
\begin{tabular}{c|ccccc}
$M_{1}$ & $X$ & $Z$ & $Z$ & $X$ & $I$\tabularnewline
$M_{2}$ & $I$ & $X$ & $Z$ & $Z$ & $X$\tabularnewline
$M_{3}$ & $X$ & $I$ & $X$ & $Z$ & $Z$\tabularnewline
$M_{4}$ & $Z$ & $X$ & $I$ & $X$ & $Z$\tabularnewline
$\bar{X}$ & $X$ & $X$ & $X$ & $X$ & $X$\tabularnewline
$\bar{Z}$ & $Z$ & $Z$ & $Z$ & $Z$ & $Z$\tabularnewline
\end{tabular}
\par\end{center}

\subsection{Extended parity-check matrix and encoder design}

In \cite{gottesman-thesis}, the parity check matrix of the five-qubit
code using binary formalism was given. Using the binary formalism as described in Section \ref{sec-2-b}, we can write the extended parity-check matrix as follows:

\begin{equation}
H_{q}=\left[\begin{array}{ccc}
\begin{array}{ccccc}
1 & 0 & 0 & 1 & 0\\
0 & 1 & 0 & 0 & 1\\
1 & 0 & 1 & 0 & 0\\
0 & 1 & 0 & 1 & 0
\end{array} & \Bigg| & \begin{array}{ccccc}
0 & 1 & 1 & 0 & 0\\
0 & 0 & 1 & 1 & 0\\
0 & 0 & 0 & 1 & 1\\
1 & 0 & 0 & 0 & 1
\end{array}\end{array}\right]
\end{equation}

For the encoder design, $H_{q}$ is converted to standard form using
Gaussian elimination. The standard form was given in \cite{gottesman-thesis}
directly; however, we describe the steps in detail. Our aim is to bring
the above parity check matrix in the standard form through Gaussian
elimination as described in Section \ref{sec:systematic}.
Applying $R_{3}\rightarrow R_{3}+R_{1}$ and $R_{4}\rightarrow R_{4}+R_{2}$

\begin{equation}
H_{q}=\left[\begin{array}{ccc}
\begin{array}{ccccc}
1 & 0 & 0 & 1 & 0\\
0 & 1 & 0 & 0 & 1\\
0 & 0 & 1 & 1 & 0\\
0 & 0 & 0 & 1 & 1
\end{array} & \Bigg| & \begin{array}{ccccc}
0 & 1 & 1 & 0 & 0\\
0 & 0 & 1 & 1 & 0\\
0 & 1 & 1 & 1 & 1\\
1 & 0 & 1 & 1 & 1
\end{array}\end{array}\right]
\end{equation}

Applying $R_{1}\rightarrow R_{1}+R_{4}$ and $R_{3}\rightarrow R_{3}+R_{4}$,
we get the standard form of the parity check matrix as

\begin{equation}
H_{s}=\left[\begin{array}{ccc}
\begin{array}{ccccc}
1 & 0 & 0 & 0 & 1\\
0 & 1 & 0 & 0 & 1\\
0 & 0 & 1 & 0 & 1\\
0 & 0 & 0 & 1 & 1
\end{array} & \Bigg| & \begin{array}{ccccc}
1 & 1 & 0 & 1 & 1\\
0 & 0 & 1 & 1 & 0\\
1 & 1 & 0 & 0 & 0\\
1 & 0 & 1 & 1 & 1
\end{array}\end{array}\right]
\end{equation}

We observe that $A_{2}=\left[\begin{array}{c}
1\\
1\\
1\\
1
\end{array}\right]$, $B=\left[\begin{array}{cccc}
1 & 1 & 0 & 1\\
0 & 0 & 1 & 1\\
1 & 1 & 0 & 0\\
1 & 0 & 1 & 1
\end{array}\right]$, and $C_{2}=\left[\begin{array}{c}
1\\
0\\
0\\
1
\end{array}\right]$

The logical operators can be evaluated using the steps mentioned in Section \ref{sec:systematic}. We get,

\begin{equation}
\overline{X}=\left[\begin{array}{ccc}
00001 & | & 10010\end{array}\right]
\end{equation}
\begin{equation}
\overline{Z}=\left[\begin{array}{ccc}
00000 & | & 11111\end{array}\right]
\end{equation}

From the standard parity-check matrix $H_s$ and the logical operators $\overline{X}$ and $\overline{Z}$, we have,

\begin{center}
\begin{tabular}{c|ccccc}
$M_{1}$ & $Y$ & $Z$ & $I$ & $Z$ & $Y$\tabularnewline
$M_{2}$ & $I$ & $X$ & $Z$ & $Z$ & $X$\tabularnewline
$M_{3}$ & $Z$ & $Z$ & $X$ & $I$ & $X$\tabularnewline
$M_{4}$ & $Z$ & $I$ & $Z$ & $Y$ & $Y$\tabularnewline
$\bar{X}$ & $Z$ & $I$ & $I$ & $Z$ & $X$\tabularnewline
$\bar{Z}$ & $Z$ & $Z$ & $Z$ & $Z$ & $Z$\tabularnewline
\end{tabular}
\par\end{center}The basis codewords for this code can be written as

\begin{equation}
|\bar{0}\rangle=\sum_{M\in S}M|00000\rangle
\end{equation}

\begin{equation}
|\bar{1}\rangle=\bar{X}|\bar{0}\rangle
\end{equation}

which gives us the encoded $|\bar{0}\rangle$ as

\begin{align}
|\bar{0}\rangle & =|00000\rangle+M_{1}|00000\rangle+M_{2}|00000\rangle+M_{3}|00000\rangle\nonumber \\
 & +M_{4}|00000\rangle+M_{1}M_{2}|00000\rangle+M_{1}M_{3}|00000\rangle\nonumber \\
 & +M_{1}M_{4}|00000\rangle+M_{2}M_{3}|00000\rangle+M_{2}M_{4}|00000\rangle\nonumber \\
 & +M_{3}M_{4}|00000\rangle+M_{1}M_{2}M_{3}|00000\rangle\nonumber \\
 & +M_{1}M_{2}M_{4}|00000\rangle+M_{1}M_{3}M_{4}|00000\rangle\nonumber \\
 & +M_{2}M_{3}M_{4}|00000\rangle+M_{1}M_{2}M_{3}M_{4}|00000\rangle\nonumber \\
 & =\frac{1}{4}(|00000\rangle+|10010\rangle+|01001\rangle+|10100\rangle+|01010\rangle\nonumber \\
 & -|11011\rangle-|00110\rangle-|11000\rangle-|11101\rangle-|00011\rangle\nonumber \\
 & -|11110\rangle-|01111\rangle-|10001\rangle-|01100\rangle-|10111\rangle\nonumber \\
 & +|00101\rangle)\label{eq:5-q-0}
\end{align}

and encoded $|\bar{1}\rangle$ as

\begin{align}
|\bar{1}\rangle & =\bar{X}|\bar{0}\rangle\nonumber \\
 & =\frac{1}{4}(-|11111\rangle-|01101\rangle-|10110\rangle-|01011\rangle-|10101\rangle\nonumber \\
 & +|00100\rangle+|11001\rangle+|00111\rangle+|00010\rangle+|11100\rangle\nonumber \\
 & +|00001\rangle+|10000\rangle+|01110\rangle+|10011\rangle+|01000\rangle\nonumber \\
 & -|11010\rangle)\label{eq:5-q-1}
\end{align}

\begin{figure}
\begin{centering}
\includegraphics[scale=0.4]{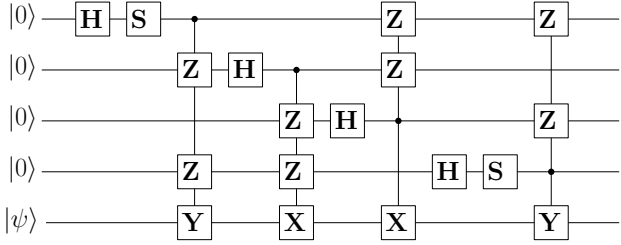}
\par\end{centering}
\caption{Encoder for the five-qubit code. \label{fig:five-qubit-encoder}}
\end{figure}

Following the procedure in \ref{sec:systematic}, we put the input qubit $|\psi\rangle$ at the $5^\mathrm{th}$ spot followed by $n-1$ qubits initialized to $|0\rangle$ state. Next, the logical operators are encoded according the Algorithm \ref{alg:algo-1}. Thereafter, the stabilizers corresponding to the rows of standard form of the parity check matrix $H_{s}$ are applied according the Algorithm \ref{alg:algo-1}. The encoder circuit thus designed is shown in Fig. \ref{fig:five-qubit-encoder}. 

\begin{figure}
\begin{centering}
\includegraphics[scale=0.4]{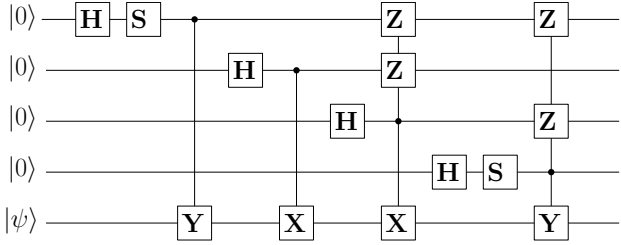}
\par\end{centering}
\caption{Modified encoder for the five-qubit code. \label{fig:five-qubit-encoder-modified}}
\end{figure}

Next, we observe that there are four $Z$  gates which are acting on state $|0\rangle$, making those $Z$  gates redundant. After removing those $Z$  gates, the modified encoding circuit is shown in Fig. \ref{fig:five-qubit-encoder-modified}.

On observing carefully, we notice that this circuit is slightly different from the encoder provided in \cite{gottesman-thesis}. In the circuit in \cite{gottesman-thesis}, there is a subtle error due to which the stabilizers don't commute. To be specific, the $H$  gate (or $H$ followed by $S$  gate) should appear just before the control dots, else the stabilizer operators don't commute. Also, the circuit in \cite{gottesman-thesis} 
 uses $Z$  gates instead of $S$ after the $H$  gates when required. However, if one intends to use $Z$  gate, one has to use controlled-$X$  followed by controlled-$Z$ 
 instead of controlled-$Y$  gates. These errors were later addressed in an errata \cite{gottesman-errata}. 

\subsection{Syndrome measurement circuit and error corrector}

The syndrome measurement circuit measures the four stabilizers using
four ancilla qubits initialized to the $|0\rangle$ state. There are
$5$ qubits and each qubit can be affected by $X$, $Y$, or $Z$
errors. So, $15$ unique error syndromes are possible, which are represented
by the final state of the ancilla qubits. The syndromes are shown
in Table \ref{tab:tab-syn-five-qubit}. 

\begin{table*}
\caption{Syndrome table for the 5-qubit code.\label{tab:tab-syn-five-qubit}}

\centering{}
\global\long
\begin{tabular}{|ccccc|cccc|c|}
\hline 
 &  &  &  &  & \multicolumn{1}{c|}{$M_{1}$} & \multicolumn{1}{c|}{$M_{2}$} & \multicolumn{1}{c|}{$M_{3}$} & $M_{4}$ & Decimal value\tabularnewline
\hline 
\hline 
$X$ & $I$ & $I$ & $I$ & $I$ & $0$ & $0$ & $0$ & $1$ & $1$\tabularnewline
\hline 
$Z$ & $I$ & $I$ & $I$ & $I$ & $1$ & $0$ & $1$ & $0$ & $10$\tabularnewline
\hline 
$Y$ & $I$ & $I$ & $I$ & $I$ & $1$ & $0$ & $1$ & $1$ & $11$\tabularnewline
\hline 
$I$ & $X$ & $I$ & $I$ & $I$ & $1$ & $0$ & $0$ & $0$ & $8$\tabularnewline
\hline 
$I$ & $Z$ & $I$ & $I$ & $I$ & $0$ & $1$ & $0$ & $1$ & $5$\tabularnewline
\hline 
$I$ & $Y$ & $I$ & $I$ & $I$ & $1$ & $1$ & $0$ & $1$ & $13$\tabularnewline
\hline 
$I$ & $I$ & $X$ & $I$ & $I$ & $1$ & $1$ & $0$ & $0$ & $12$\tabularnewline
\hline 
$I$ & $I$ & $Z$ & $I$ & $I$ & $0$ & $0$ & $1$ & $0$ & $2$\tabularnewline
\hline 
$I$ & $I$ & $Y$ & $I$ & $I$ & $1$ & $1$ & $1$ & $0$ & $14$\tabularnewline
\hline 
$I$ & $I$ & $I$ & $X$ & $I$ & $0$ & $1$ & $1$ & $0$ & $6$\tabularnewline
\hline 
$I$ & $I$ & $I$ & $Z$ & $I$ & $1$ & $0$ & $0$ & $1$ & $9$\tabularnewline
\hline 
$I$ & $I$ & $I$ & $Y$ & $I$ & $1$ & $1$ & $1$ & $1$ & $15$\tabularnewline
\hline 
$I$ & $I$ & $I$ & $I$ & $X$ & $0$ & $0$ & $1$ & $1$ & $3$\tabularnewline
\hline 
$I$ & $I$ & $I$ & $I$ & $Z$ & $0$ & $1$ & $0$ & $0$ & $4$\tabularnewline
\hline 
$I$ & $I$ & $I$ & $I$ & $Y$ & $0$ & $1$ & $1$ & $1$ & $7$\tabularnewline
\hline 
$I$ & $I$ & $I$ & $I$ & $I$ & $0$ & $0$ & $0$ & $0$ & $0$\tabularnewline
\hline 
\end{tabular}
\end{table*}

\begin{figure}
\begin{centering}
\includegraphics[scale=0.35]{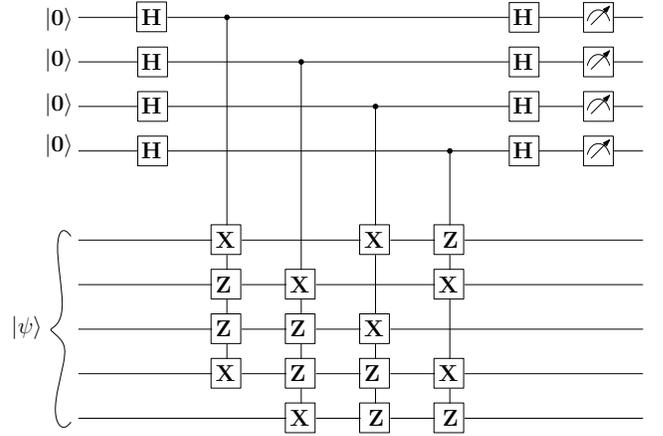}
\par\end{centering}
\caption{Syndrome measurement circuit for 5-qubit code. \label{fig:syn-measure-five-qubit}}
\end{figure}

The syndrome measurement circuit is shown in Fig. \ref{fig:syn-measure-five-qubit}.
Depending on the syndrome, appropriate error correction can be performed
by using suitable $X$, $Z$, or $Y$ gate on the appropriate qubit.

\subsection{Evaluation of the output state of the encoder circuit}

A good exercise would be to evaluate the output state of the encoder circuit and verify if it matches with $|\overline{0}\rangle$ and $|\overline{1}\rangle$ states in equations \ref{eq:5-q-0} and \ref{eq:5-q-1}. 

The initial state $\psi_{0}$ when the fifth qubit is set to $|0\rangle$ is $\psi_{0}=|00000\rangle$. 

\textbf{Step A:} Applying $H$ gate on qubit $1$ followed by $S$, we have

\begin{align}
|\psi_1\rangle= & \frac{1}{\sqrt{2}} (|00000\rangle+i|10000\rangle)
\end{align}

\textbf{Step B:} Applying $M_1$ controlled at qubit $1$ we have,

\begin{align}
|\psi_2\rangle= & \frac{1}{\sqrt{2}} (|00000\rangle+(i\cdot i)|10001\rangle)
\nonumber \\
=& \frac{1}{\sqrt{2}} (|00000\rangle -|10001\rangle)
\end{align}

\textbf{Step C:} Applying $H$ gate on qubit $2$, we have

\begin{align}
|\psi_3\rangle= & \frac{1}{2} (|00000\rangle+|01000\rangle-|10001\rangle-|11001\rangle)
\end{align}

\textbf{Step D:} Applying $M_2$ controlled at qubit $2$ we have,

\begin{align}
|\psi_4\rangle= & \frac{1}{2} (|00000\rangle+|01001\rangle-|10001\rangle-|11000\rangle)
\end{align}

\textbf{Step E:} Applying $H$ gate on qubit $3$, we have

\begin{align}
|\psi_5\rangle= & \frac{1}{2\sqrt{2}} (|00000\rangle+|00100\rangle+|01001\rangle+|01101\rangle\nonumber\\
& -|10001\rangle-|10101\rangle-|11000\rangle-|11100\rangle)
\end{align}

\textbf{Step F:} Applying $M_3$ controlled at qubit $3$, we have

\begin{align}
|\psi_6\rangle= & \frac{1}{2\sqrt{2}} (|00000\rangle+|00101\rangle+|01001\rangle-|01100\rangle\nonumber\\
& -|10001\rangle+|10100\rangle-|11000\rangle-|11101\rangle)
\end{align}

\textbf{Step G:} Applying $H$ gate on qubit $4$ followed by $S$ gate, we have

\begin{align}
|\psi_7\rangle & =\frac{1}{4} (|00000\rangle+i|00010\rangle+|00101\rangle+i|00111\rangle\nonumber\\
& +|01001\rangle+i|01011\rangle-|01100\rangle-i|01110\rangle\nonumber\\
& -|10001\rangle-i|10011\rangle+|10100\rangle+i|10110\rangle\nonumber\\
& -|11000\rangle-i|11010\rangle-|11101\rangle-i|11111\rangle)
\end{align}

\textbf{Step H:} Applying $M_4$ controlled at qubit $4$, we have

\begin{align}
|\psi_8\rangle & =\frac{1}{4} (|00000\rangle+i\cdot i|00011\rangle+|00101\rangle\nonumber\\
& +i(-i\cdot -1)|00110\rangle
+|01001\rangle+i(-i)|01010\rangle\nonumber\\
& -|01100\rangle-i(-1\cdot i)|01111\rangle-|10001\rangle\nonumber\\
&-i(-1\cdot -i)|10010\rangle+|10100\rangle+i(i\cdot -1\cdot -1)|10111\rangle\nonumber\\
& -|11000\rangle-i(-1\cdot i)|11011\rangle-|11101\rangle\nonumber\\
& -i(-1\cdot -1\cdot -i)|11110\rangle)\\ 
& =\frac{1}{4} (|00000\rangle-|00011\rangle+|00101\rangle-|00110\rangle\nonumber\\
& +|01001\rangle+|01010\rangle-|01100\rangle-|01111\rangle\nonumber\\
& -|10001\rangle+|10010\rangle+|10100\rangle-|10111\rangle\nonumber\\
& -|11000\rangle-|11011\rangle-|11101\rangle-|11110\rangle)\nonumber\\
\end{align}

We observe that $|\psi_{8}\rangle$ matches with state $|\overline{0}\rangle$ in equation \ref{eq:5-q-0}. 

Now, we will verify state $|\overline{1}\rangle$. The initial state $\psi_{0}$ when the fifth qubit is set to $|1\rangle$ is $\psi_{0}=|00001\rangle$. 

\textbf{Step A:} Applying $H$ gate on qubit $1$ followed by $S$, we have

\begin{align}
|\psi_1\rangle= & \frac{1}{\sqrt{2}} (|00001\rangle+i|10001\rangle)
\end{align}

\textbf{Step B:} Applying $M_1$ controlled at qubit $1$ we have,

\begin{align}
|\psi_2\rangle= & \frac{1}{\sqrt{2}} (|00001\rangle+(i\cdot -i)|10000\rangle)
\nonumber \\
=& \frac{1}{\sqrt{2}} (|00001\rangle +|10000\rangle)
\end{align}

\textbf{Step C:} Applying $H$ gate on qubit $2$, we have

\begin{align}
|\psi_3\rangle= & \frac{1}{2} (|00001\rangle+|01001\rangle+|10000\rangle+|11000\rangle)
\end{align}

\textbf{Step D:} Applying $M_2$ controlled at qubit $2$ we have,

\begin{align}
|\psi_4\rangle= & \frac{1}{2} (|00001\rangle+|01000\rangle+|10000\rangle+|11001\rangle)
\end{align}

\textbf{Step E:} Applying $H$ gate on qubit $3$, we have

\begin{align}
|\psi_5\rangle= & \frac{1}{2\sqrt{2}} (|00001\rangle+|00101\rangle+|01000\rangle+|01100\rangle\nonumber\\
& +|10000\rangle+|10100\rangle+|11001\rangle+|11101\rangle)
\end{align}

\textbf{Step F:} Applying $M_3$ controlled at qubit $3$, we have
\begin{align}
|\psi_6\rangle= & \frac{1}{2\sqrt{2}} (|00001\rangle+|00100\rangle+|01000\rangle-|01101\rangle\nonumber\\
& +|10000\rangle-|10101\rangle+|11001\rangle+|11100\rangle)
\end{align}

\textbf{Step G:} Applying $H$ gate on qubit $4$ followed by $S$ gate, we have
\begin{align}
|\psi_7\rangle & =\frac{1}{4} (|00001\rangle+i|00011\rangle+|00100\rangle+i|00110\rangle\nonumber\\
& +|01000\rangle+i|01010\rangle-|01101\rangle-i|01111\rangle\nonumber\\
& +|10000\rangle+i|10010\rangle-|10101\rangle-i|10111\rangle\nonumber\\
& +|11001\rangle+i|11011\rangle+|11100\rangle+i|11110\rangle)
\end{align}

\textbf{Step H:} Applying $M_4$ controlled at qubit $4$, we have
\begin{align}
|\psi_8\rangle & =\frac{1}{4} (|00001\rangle+i\cdot (-i)|00010\rangle+|00100\rangle\nonumber\\
& +i(-1\cdot i)|00111\rangle
+|01000\rangle+i(i)|01011\rangle\nonumber\\
& -|01101\rangle-i(-1\cdot -i)|01110\rangle+|10000\rangle\nonumber\\
&+i(-1\cdot i)|10011\rangle-|10101\rangle-i(-1\cdot -1\cdot -i)|10110\rangle\nonumber\\
& +|11001\rangle+i(-1\cdot -i)|11010\rangle+|11100\rangle\nonumber\\
& +i(-1\cdot -1\cdot i)|11111\rangle)\\ 
& =\frac{1}{4} (|00001\rangle+|00010\rangle+|00100\rangle+|00111\rangle\nonumber\\
& +|01000\rangle-|01011\rangle-|01101\rangle+|01110\rangle\nonumber\\
& +|10000\rangle+|10011\rangle-|10101\rangle-|10110\rangle\nonumber\\
& +|11001\rangle-|11010\rangle+|11100\rangle-|11111\rangle)\nonumber\\
\end{align}

We observe that $|\psi_{8}\rangle$ matches with state $|\overline{1}\rangle$ in equation \ref{eq:5-q-1}. 

\section{Classical $[7,4,3]$ Hamming code and Steane code}

Hamming codes \cite{hamming} are linear error correcting codes which
have the property that they can detect 1- and 2-bit errors, and can
correct 1-bit errors. The $[7,4,3]$ Hamming code was introduced
by Hamming. It encodes $4$ bits of data into $7$ bits, such that
the $3$ parity bits provide the ability to detect and correct single
bit errors. The generator matrix $G$ and the parity check matrix
$H$ of the Hamming code are given as,

\begin{align}
G=\left[\begin{array}{c}
1000110\\
0100101\\
0010011\\
0001111
\end{array}\right],H & =\left[\begin{array}{c}
1101100\\
1011010\\
0111001
\end{array}\right]
\end{align}

\subsection{Steane code as the quantum analog of classical Hamming code}

Steane code \cite{steane-2} is a CSS code which uses the Hamming
$[7,4,3]$ code and the dual of the Hamming code, i.e., the $[7,3,4]$
code to correct bit flip and phase flip errors respectively. The $[7,4,3]$
Hamming code contains its dual, and thus can be used in the CSS framework
to obtain a quantum ECC. One logical qubit is encoded into seven physical
qubits, thus enabling the Steane code to detect and correct a single
qubit error. In stabilizer framework, the Steane code is represented
by six generators as shown below:
\begin{center}
\begin{tabular}{c|ccccccc}
$M_{1}$  & $X$  & $X$  & $X$  & $X$  & $I$  & $I$  & $I$\tabularnewline
$M_{2}$  & $X$  & $X$  & $I$  & $I$  & $X$  & $X$  & $I$\tabularnewline
$M_{3}$  & $X$  & $I$  & $X$  & $I$  & $X$  & $I$  & $X$\tabularnewline
$M_{4}$  & $Z$  & $Z$  & $Z$  & $Z$  & $I$  & $I$  & $I$\tabularnewline
$M_{5}$  & $Z$  & $Z$  & $I$  & $I$  & $Z$  & $Z$  & $I$\tabularnewline
$M_{6}$  & $Z$  & $I$  & $Z$  & $I$  & $Z$  & $I$  & $Z$\tabularnewline
\end{tabular}. 
\par\end{center}

Each of the above generators is a tensor product of $7$ Pauli matrices.
It should however be noted that tensor product symbols $\otimes$
are often ommited for brevity. The logical operators are $X_{L}=XXXXXXX$
and $Z_{L}=ZZZZZZZ$. Thus, the two codewords for the Steane code
are,

\begin{align}
|0\rangle_{L} & =\frac{1}{2\sqrt{2}}(|0000000\rangle+|1111000\rangle+|1100110\rangle+|1010101\rangle\nonumber \\
 & +|0011110\rangle+|0101101\rangle+|0110011\rangle+|1001011\rangle)\\
|1\rangle_{L} & =X_{L}|0\rangle\nonumber \\
 & =\frac{1}{2\sqrt{2}}(|0000111\rangle+|1111111\rangle+|1100001\rangle+|1010010\rangle\nonumber \\
 & +|0011001\rangle+|0101010\rangle+|0110100\rangle+|1001100\rangle)
\end{align}

\subsection{Encoder designed by converting extended parity check matrix to standard
form using Gaussian elimination}

We have the parity check matrix and generator matrix for (7,4) Hamming
code as follows:

\begin{align}
H & =\left[\begin{array}{c}
1101100\\
1011010\\
0111001
\end{array}\right],G=\left[\begin{array}{c}
1000110\\
0100101\\
0010011\\
0001111
\end{array}\right]
\end{align}

We can verify that $H$ is contained in $G$. Thus, it satisfies the
dual-containing criterion for construction of CSS codes. In the binary
formalism, the parity check matrix for the augmented parity check
matrix can be written as

\begin{equation}
H_{q}=\left[\begin{array}{ccc}
\begin{array}{c}
1101100\\
1011010\\
0111001\\
0000000\\
0000000\\
0000000
\end{array} & \Bigg| & \begin{array}{c}
0000000\\
0000000\\
0000000\\
1101100\\
1011010\\
0111011
\end{array}\end{array}\right]
\end{equation}

Our aim is to transform the above parity check matrix to the standard
form as described in Section \ref{sec:systematic}.  
First, some columns are swapped, which is equivalent to swapping
qubit positions. The columns (or equivalently the qubit positions)
are swapped in following order $1\leftarrow5,\,2\leftarrow6,\,3\leftarrow7,\,4\leftarrow1,\,5\leftarrow2,\,6\leftarrow4,\,7\leftarrow3$.
This gives us the new augmented $H$ matrix
\begin{equation}
H_{q}=\left[\begin{array}{ccc}
\begin{array}{c}
1001110\\
0101011\\
0010111\\
0000000\\
0000000\\
0000000
\end{array} & \Bigg| & \begin{array}{c}
0000000\\
0000000\\
0000000\\
1001110\\
0101011\\
0010111
\end{array}\end{array}\right]
\end{equation}

Performing the operation $R_{5}\rightarrow R_{5}+R_{4}$

\begin{equation}
H_{q}=\left[\begin{array}{ccc}
\begin{array}{c}
1001110\\
0101011\\
0010111\\
0000000\\
0000000\\
0000000
\end{array} & \Bigg| & \begin{array}{c}
0000000\\
0000000\\
0000000\\
1001110\\
1100101\\
0010111
\end{array}\end{array}\right]
\end{equation}

Performing the operation $R_{6}\rightarrow R_{6}+R_{5}$

\begin{equation}
H_{q}=\left[\begin{array}{ccc}
\begin{array}{c}
1001110\\
0101011\\
0010111\\
0000000\\
0000000\\
0000000
\end{array} & \Bigg| & \begin{array}{c}
0000000\\
0000000\\
0000000\\
1001110\\
1100101\\
1110010
\end{array}\end{array}\right]
\end{equation}

Performing the operation $R_{4}\rightarrow R_{4}+R_{5}$

\begin{equation}
H_{q}=\left[\begin{array}{ccc}
\begin{array}{c}
1001110\\
0101011\\
0010111\\
0000000\\
0000000\\
0000000
\end{array} & \Bigg| & \begin{array}{c}
0000000\\
0000000\\
0000000\\
0101011\\
1100101\\
1110010
\end{array}\end{array}\right]
\end{equation}

Performing the operation $R_{4}\rightarrow R_{4}+R_{6}$ we get the
standard form $H_{s}$ as

\begin{equation}
H_{s}=\left[\begin{array}{ccc}
\begin{array}{c}
1001110\\
0101011\\
0010111\\
0000000\\
0000000\\
0000000
\end{array} & \Bigg| & \begin{array}{c}
0000000\\
0000000\\
0000000\\
1011001\\
1100101\\
1110010
\end{array}\end{array}\right]
\end{equation}

We have the following from $H_{s}$.
\begin{align}
I_{1} & =I_{2}=\left[\begin{array}{ccc}
1 & 0 & 0\\
0 & 1 & 0\\
0 & 0 & 1
\end{array}\right],A_{1=}\left[\begin{array}{ccc}
1 & 1 & 1\\
1 & 0 & 1\\
0 & 1 & 1
\end{array}\right],\nonumber \\
A_{2} & =\left[\begin{array}{c}
0\\
1\\
1
\end{array}\right],B=C_{1}=0_{3\times3},C_{2}=0_{3\times1},\nonumber \\
D & =\left[\begin{array}{ccc}
1 & 0 & 1\\
1 & 1 & 0\\
1 & 1 & 1
\end{array}\right],E=\left[\begin{array}{c}
1\\
1\\
0
\end{array}\right].
\end{align}

The stabilizers of the code can be written as
\begin{center}
\begin{tabular}{c|ccccccc}
$M_{1}$ & $X$ & $I$ & $I$ & $X$ & $X$ & $X$ & $I$\tabularnewline
$M_{2}$ & $I$ & $X$ & $I$ & $X$ & $I$ & $X$ & $X$\tabularnewline
$M_{3}$ & $I$ & $I$ & $X$ & $I$ & $X$ & $X$ & $X$\tabularnewline
$M_{4}$ & $Z$ & $I$ & $I$ & $Z$ & $Z$ & $Z$ & $I$\tabularnewline
$M_{5}$ & $I$ & $Z$ & $I$ & $Z$ & $I$ & $Z$ & $Z$\tabularnewline
$M_{6}$ & $I$ & $I$ & $Z$ & $I$ & $Z$ & $Z$ & $Z$\tabularnewline
\end{tabular}
\par\end{center}

The logical operators can be evaluated as described in Section \ref{sec:systematic}, producing
\begin{equation}
\overline{X}=\left[\begin{array}{ccc}
0001101 & | & 0000000\end{array}\right]
\end{equation}

\begin{equation}
\overline{Z}=\left[\begin{array}{ccc}
0000000 & | & 0110001\end{array}\right]
\end{equation}

\begin{figure}
\begin{centering}
\includegraphics[scale=0.32]{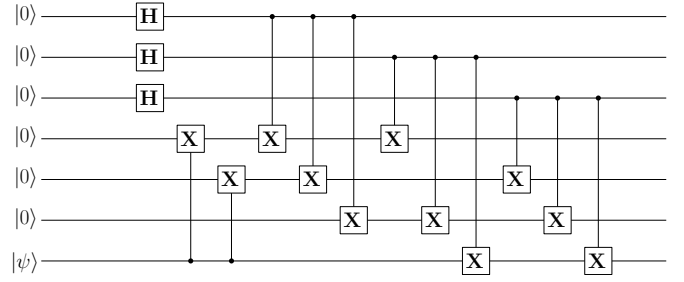}
\par\end{centering}
\caption{Encoder for the Steane code. \label{fig:encoder-ge-1}}
\end{figure}

The encoding circuit can be generated from $\overline{X}$ and $H_{s}$ by applying Algorithm \ref{alg:algo-1}. The qubit to be encoded is placed at the $7^\mathrm{th}$ position, followed by $6$ qubits initialized to $|0\rangle$. First, the $\overline{X}|0000000\rangle$ state is obtained by applying $\overline{X}$ conditioned on the last qubit. Applying Algorithm \ref{alg:algo-1}, the encoder circuit thus obtained is shown in Fig. \ref{fig:encoder-ge-1}.

\subsection{Syndrome measurement circuit and error corrector}

The syndrome measurement circuit measures all the six stabilizers
using six ancilla qubits. The syndromes are unique as shown in Table
\ref{tab:tab-syn-steane}. Each qubit in the 7-qubit Steane code can
be affected by three kind of errors, namely $X$, $Y$, and $Z$ errors.
So, there are $21$ different types of single qubit errors possible,
each of which gives a different syndrome as shown in Table \ref{tab:tab-syn-steane}.
The $M_{1}$-$M_{6}$ values in the table can be explained by the
following example. Let us take the fifth row of the table for example,
i.e., $IZIIIII$, which implies that a $Z$ error has occurred on the
second qubit. It is easy to observe that $IZIIIII$ anti-commutes with
$M_{1}$ and $M_{2}$, while it commutes with $M_{3},$$M_{4}$, $M_{5}$,
and $M_{6}.$ Thus, we get a syndrome of $110000$. It can be observed
that each syndrome is unique as shown in Table \ref{tab:tab-syn-steane}.
Since this code uses only $21$ different syndromes for various single
qubit errors, the rest of the syndromes are unused, unlike the 5-qubit
perfect code where all the syndromes are used.

\begin{table*}
\caption{Syndrome table for the Steane code.\label{tab:tab-syn-steane}}

\centering{}
\global\long
\begin{tabular}{|ccccccc|cccccc|c|}
\hline 
 &  &  &  &  &  &  & \multicolumn{1}{c|}{$M_{1}$} & \multicolumn{1}{c|}{$M_{2}$} & \multicolumn{1}{c|}{$M_{3}$} & \multicolumn{1}{c|}{$M_{4}$} & \multicolumn{1}{c}{$M_{5}$} & $M_{6}$  & Decimal value\tabularnewline
\hline 
\hline 
$X$  & $I$  & $I$  & $I$  & $I$  & $I$  & $I$  & $0$  & $0$  & $0$  & $1$  & $0$  & $0$  & $4$\tabularnewline
\hline 
$Z$  & $I$  & $I$  & $I$  & $I$  & $I$  & $I$  & $1$  & $0$  & $0$  & $0$  & $0$  & $0$  & $32$\tabularnewline
\hline 
$Y$  & $I$  & $I$  & $I$  & $I$  & $I$  & $I$  & $1$  & $0$  & $0$  & $1$  & $0$  & $0$  & $36$\tabularnewline
\hline 
$I$  & $X$  & $I$  & $I$  & $I$  & $I$  & $I$  & $0$  & $0$  & $0$  & $0$  & $1$  & $0$  & $2$\tabularnewline
\hline 
$I$  & $Z$  & $I$  & $I$  & $I$  & $I$  & $I$  & $0$  & $1$  & $0$  & $0$  & $0$  & $0$  & $16$\tabularnewline
\hline 
$I$  & $Y$  & $I$  & $I$  & $I$  & $I$  & $I$  & $0$  & $1$  & $0$  & $0$  & $1$  & $0$  & $18$\tabularnewline
\hline 
$I$  & $I$  & $X$  & $I$  & $I$  & $I$  & $I$  & $0$  & $0$  & $0$  & $0$  & $0$  & $1$  & $1$\tabularnewline
\hline 
$I$  & $I$  & $Z$  & $I$  & $I$  & $I$  & $I$  & $0$  & $0$  & $1$  & $0$  & $0$  & $0$  & $8$\tabularnewline
\hline 
$I$  & $I$  & $Y$  & $I$  & $I$  & $I$  & $I$  & $0$  & $0$  & $1$  & $0$  & $0$  & $1$  & $9$\tabularnewline
\hline 
$I$  & $I$  & $I$  & $X$  & $I$  & $I$  & $I$  & $0$  & $0$  & $0$  & $1$  & $1$  & $0$  & $6$\tabularnewline
\hline 
$I$  & $I$  & $I$  & $Z$  & $I$  & $I$  & $I$  & $1$  & $1$  & $0$  & $0$  & $0$  & $0$  & $48$\tabularnewline
\hline 
$I$  & $I$  & $I$  & $Y$  & $I$  & $I$  & $I$  & $1$  & $1$  & $0$  & $1$  & $1$  & $0$  & $54$\tabularnewline
\hline 
$I$  & $I$  & $I$  & $I$  & $X$  & $I$  & $I$  & $0$  & $0$  & $0$  & $1$  & $0$  & $1$  & $5$\tabularnewline
\hline 
$I$  & $I$  & $I$  & $I$  & $Z$  & $I$  & $I$  & $1$  & $0$  & $1$  & $0$  & $0$  & $0$  & $40$\tabularnewline
\hline 
$I$  & $I$  & $I$  & $I$  & $Y$  & $I$  & $I$  & $1$  & $0$  & $1$  & $1$  & $0$  & $1$  & $45$\tabularnewline
\hline 
$I$  & $I$  & $I$  & $I$  & $I$  & $X$  & $I$  & $0$  & $0$  & $0$  & $1$  & $1$  & $1$  & $7$\tabularnewline
\hline 
$I$  & $I$  & $I$  & $I$  & $I$  & $Z$  & $I$  & $1$  & $1$  & $1$  & $0$  & $0$  & $0$  & $56$\tabularnewline
\hline 
$I$  & $I$  & $I$  & $I$  & $I$  & $Y$  & $I$  & $1$  & $1$  & $1$  & $1$  & $1$  & $1$  & $63$\tabularnewline
\hline 
$I$  & $I$  & $I$  & $I$  & $I$  & $I$  & $X$  & $0$  & $0$  & $0$  & $0$  & $1$  & $1$  & $3$\tabularnewline
\hline 
$I$  & $I$  & $I$  & $I$  & $I$  & $I$  & $Z$  & $0$  & $1$  & $1$  & $0$  & $0$  & $0$  & $24$\tabularnewline
\hline 
$I$  & $I$  & $I$  & $I$  & $I$  & $I$  & $Y$  & $0$  & $1$  & $1$  & $0$  & $1$  & $1$  & $27$\tabularnewline
\hline 
$I$  & $I$  & $I$  & $I$  & $I$  & $I$  & $I$  & $0$  & $0$  & $0$  & $0$  & $0$  & $0$  & $0$\tabularnewline
\hline 
\end{tabular}
\end{table*}

\begin{figure}
\begin{centering}
\includegraphics[scale=0.35]{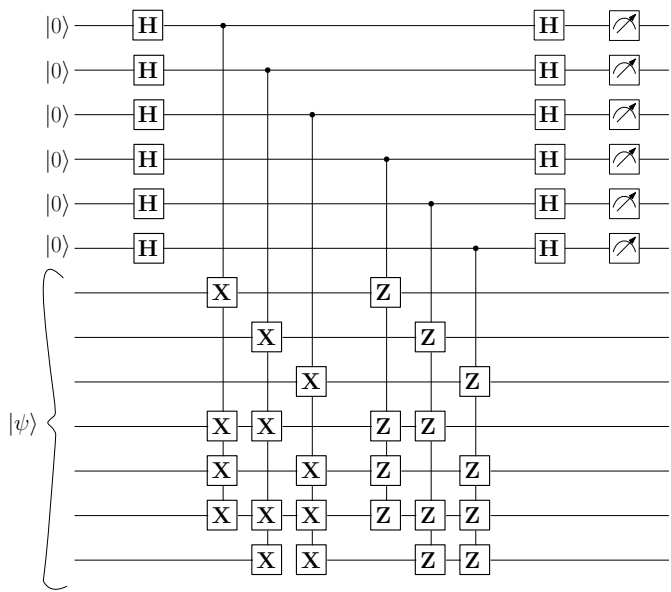} 
\par\end{centering}
\caption{Syndrome measurement circuit for Steane code. \label{fig:syn-measure-steane}}
\end{figure}

The syndrome measurement circuit is shown in Fig. \ref{fig:syn-measure-steane}.
Six ancilla qubits are used to measure each of the six stabilizers.
Measurement of the ancilla qubits gives the syndrome. Depending on
the syndrome, appropriate error correction can be performed by using
suitable $X$, $Z$, or $Y$ gate on the appropriate qubit. A syndrome
measurement of $000000$ implies that no error has occurred. It should
also be noted that any $6$ bit syndrome other than the syndromes
mentioned in Table \ref{tab:tab-syn-steane} implies the occurrence
of more than a single qubit error, which cannot be corrected using
the Steane code. 

\section{Nearest neighbour compliant circuits for the five-qubit code}

\begin{figure}
\begin{centering}
\includegraphics[scale=1]{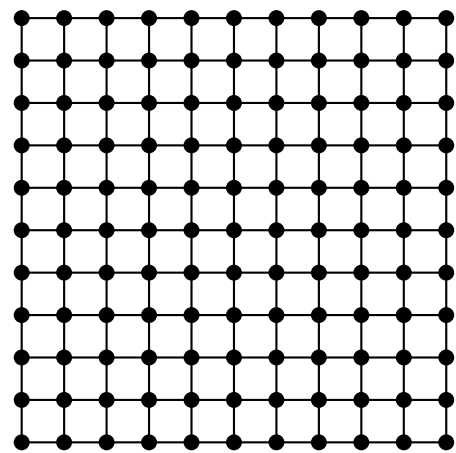} 
\par\end{centering}
\caption{2-D array of qubits (represented by black dots). The qubits can only interact with their nearest neighbours. The qubits on corners and edges can interact with 2 and 3 qubits respectively. The rest of the qubits can interact with 4 qubits each. \label{fig:2D-array}}
\end{figure}

In the circuits we discussed in the previous sections, we assume that any particular qubit can interact with any other qubit. This implies that there can be a 2-qubit gate between any two arbitrary qubits. However, it is not possible to do so in real quantum computing systems where qubits can only interact with their nearest neighbours \cite{ding}. For example, in a 2-D array of qubits in Fig. \ref{fig:2D-array}, the qubits at the corners and edges can interact with 2 or 3 qubits, while the rest of the qubits can interact with their 4 closest neighbors. 

\begin{figure}
\begin{centering}
\includegraphics[scale=0.8]{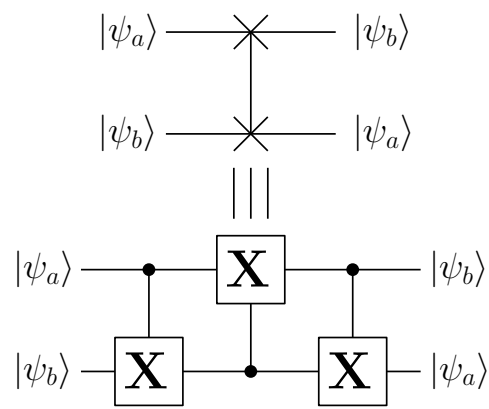} 
\par\end{centering}
\caption{Symbol of a swap gate (top). A swap gate circuit implemented using 3 CNOT gates (bottom). \label{fig:swap-gate}}
\end{figure}

To design a circuit which is nearest neighbour compliant, we need to use swap gates to bring the qubits adjacent to each other \cite{ding}. It should be noted that the qubits are not moved physically. Their states are swapped which is equivalent to moving them to adjacent positions without doing  it physically. A swap gate requires 3 CNOT gates. Thus, it is important to position the qubits and perform the operations in such a way that the number of swap gates is minimized. A swap gate is shown in Fig. \ref{fig:swap-gate}.

A nearest neighbour compliant circuit for the five-qubit encoder is shown in Fig. \ref{fig:five-qubit-encoder-NN}. The initial qubit position is shown at the top. Three swap gates are required to implement the circuit.

\begin{figure*}
\begin{centering}
\includegraphics[scale=0.45]{five-qubit-encoder-NN} 
\par\end{centering}
\caption{Nearest neighbour compliant circuit for the five-qubit encoder. The positions of the qubits in a 2-D array is shown at the top. The circuit requires 3 swap gates. \label{fig:five-qubit-encoder-NN}}
\end{figure*}

We also designed a nearest neighbour compliant circuit for the syndrome measurement circuit for the five-qubit code as shown in Fig. \ref{fig:five-qubit-measurement-NN}. The initial qubit configuration in the 2-D array is shown at the top. Eight swap gates are required for the circuit, which is equivalent to 24 CNOT gates.

\begin{figure*}
\begin{centering}
\includegraphics[scale=0.4]{five-qubit-measurement-NN} 
\par\end{centering}
\caption{Nearest neighbour compliant circuit for the five-qubit syndrome measurement circuit. The positions of the qubits in a 2-D array is shown at the top. The circuit requires 8 swap gates. \label{fig:five-qubit-measurement-NN}}
\end{figure*}

Similar to the five-qubit code, nearest neighbour compliant circuit can be designed for the Steane code encoder and and syndrome measurement circuit as well. Due to space constraints, this is not addressed in the paper.

\section{Results}

The combined encoder and decoder circuits were simulated using IBM Qiskit.
Errors were introduced at different positions to test for correctability.
Syndromes were found to match with Tables I and II for five-qubit and
Steane code, respectively.

Another important parameter to measure the efficiency of the quantum
circuits is the number of single and multiple qubit gates used by
the quantum circuits. We list the number of gates used in the quantum circuits presented in this paper in
Table \ref{tab:Resource-utilization}. 

\begin{table*}

\caption{Resource utilization summary for the various designed quantum circuits
in terms of number of gates used. \label{tab:Resource-utilization}}

\begin{centering}
\global\long\def\arraystretch{1.5}%
\begin{tabular}{|>{\centering}p{1.8cm}|>{\centering}p{1.6cm}|>{\centering}p{1.6cm}|>{\centering}p{1.6cm}|>{\centering}p{1.6cm}|}
\hline 
Parameters & Five qubit encoder & Five qubit syndrome measurement & Steane code encoder & Steane code syndrome measurement
\tabularnewline
\hline 
\hline 
H gate & 4 & 8 & 3 & 12 \tabularnewline
\hline 
S gate & 2 & 0 & 0 & 0 \tabularnewline
\hline 
Controlled X & 2 & 8 & 11 & 12 \tabularnewline
\hline 
Controlled Y & 2 & 0 & 0 & 0 \tabularnewline
\hline 
Controlled Z & 4 & 8 & 0 & 12 \tabularnewline
\hline 
\end{tabular}
\par\end{centering}
\end{table*}

\section{Conclusions}

In this paper, we provided a detailed procedure for the construction of encoding and decoding circuits for stabilizer codes. We started with Shor's 9-qubit code and analyzed the code in stabilizer formalism and then described an algorithm to generate encoding and decoding circuits of a general stabilizer code. We also provided nearest neighbour compliant circuits for the five-qubit code. Future work should be directed towards design of quantum circuits for more complex error correcting codes such as BCH codes, LDPC, and polar codes.

\end{document}